\documentclass[10pt]{article}
\usepackage{amsthm,amssymb,amsmath,amsfonts,setspace,ctable,lineno}
\usepackage{natbib}
\usepackage{graphicx,epsf}
\usepackage{epstopdf,lscape}
\usepackage[margin=1in]{geometry}

\numberwithin{equation}{section}
\theoremstyle{plain}
\newtheorem{thm}{Theorem}[section]

\newtheorem{pcp}{Principle}[section]

\begin{document}

\renewcommand{\baselinestretch}{1.2}


\renewcommand{\thefootnote}{}
$\ $\par


\fontsize{10.95}{14pt plus.8pt minus .6pt}\selectfont
\vspace{0.8pc}
\centerline{\large\bf Subgroup Mixable Inference in Personalized Medicine,} 
\centerline{\large \bf with an Application to Time-to-Event Outcomes}

\vspace{.4cm}
\centerline{Ying Ding}
\centerline{\it University of Pittsburgh}
\vspace{.4cm}
\centerline{Hui-Min Lin}
\centerline{\it University of Pittsburgh}
\vspace{.4cm}
\centerline{Jason C. Hsu}
\centerline{\it Eli Lilly and Company \& The Ohio State University}
\vspace{.55cm}
\fontsize{9}{11.5pt plus.8pt minus .6pt}\selectfont


\noindent{\bf ABSTRACT. Measuring treatment efficacy in mixture of subgroups from a randomized clinical trial is a fundamental problem in personalized medicine development, in deciding whether to treat the entire
  patient population or to target a subgroup. We show that some commonly used efficacy measures are not suitable
  for a mixture population. We also show that, while it is important to adjust for imbalance in
  the data using least squares means (LSmeans) (not marginal means) estimation,
  the current practice of applying LSmeans to directly estimate the efficacy in a mixture population for any type of outcome is inappropriate.
  Proposing a new principle called {\em subgroup mixable estimation},
  we establish the logical relationship among parameters that represent efficacy and develop a general inference procedure to confidently infer efficacy in
  subgroups and their mixtures.
Using oncology studies with time-to-event outcomes as an example, we show that Hazard Ratio is not suitable for measuring
  efficacy in a mixture population, and provide alternative efficacy measures with a valid inference procedure.}

\vspace{9pt}


\fontsize{10.95}{14pt plus.8pt minus .6pt}\selectfont

\section{Two Motivating Issues, with An Example} \label{sec:intro}

In personalized medicine (or equivalently, tailored therapeutics), we are concerned with finding whether there are subgroups of an overall patient population that exhibit a differential response to treatment. Any subgroup with a significantly better response to treatment could be identified for tailoring with appropriate labeling language and reimbursement considerations in the market. Conversely, subgroups with a worse response to treatment could be appropriately contraindicated in labeling. The best known example of a drug targeting a subgroup of patients is Herceptin for breast cancer patients with HER2/neu over-expression \citep{Herceptin2005}. More recent examples of such drugs include Xalkori for non-small cell lung cancer patients with ALK transolocation \citep{Xalkori2011}, and Zelboraf for skin cancer patients with BRAF mutation \citep{Zelboraf2011}.

In a randomized clinical trial (RCT), usually there is a treatment arm and a control arm (e.g., placebo or standard of care).
  The ``relative effect'' between the treatment and control is referred to as ``treatment efficacy''.
  Measuring efficacy in a mixture population from a RCT is a fundamental problem in
  personalized medicine development.
The patient population is thought of as a mixture of two or more
  subgroups that might derive differential efficacy from a treatment and
  an important decision to make is which subgroup or {\em combination}
  of subgroups of patients should the treatment target for. For example,
  for a biomarker that separates the population into two groups, denoted as $g^+$ and $g^-$,
  one has to decide whether to target $g^+$ only or $\{g^+,g^-\}$ combined.

In oncology, the primary endpoint is usually a time-to-event outcome,
e.g., Overall Survival (OS) or Progression Free Survival (PFS), and
  the most popular efficacy measure has been Hazard Ratio (HR) between treatment and control from a Cox Proportional Hazards (PH) model \citep{Cox1972}.
However, when subgroups exist, it is inappropriate to use HR to measure the treatment efficacy in subgroups and their mixtures,
as we illustrate in below with a recent published phase 2 oncology study.

\subsection{A motivating example} \label{subsec:ex}

\begin{figure}[tbp] 
  \begin{center}
  \includegraphics[width=0.6\textwidth]{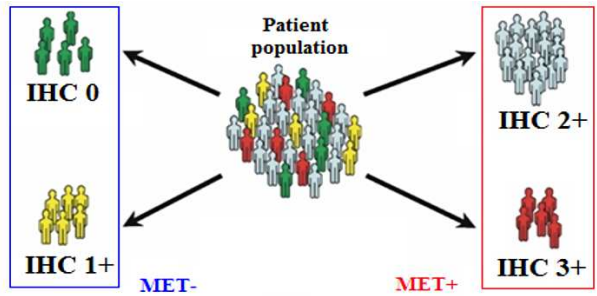}
  \caption{The plot illustrating the MET$^-$ and MET$^+$ subgroups determined by IHC in \cite{Spigel2013}.}
  \label{fig:IHCPlot}
  \end{center}
\end{figure}

\cite{Spigel2013} described a randomized phase 2 oncology study for patients with advanced non-small-cell lung cancer.
The study compared a dual treatment to a single treatment to test whether the dual treatment was more efficacious.
Two survival outcomes (PFS and OS) have been evaluated for a mixture population of patients with different MET expression levels, measured by immunohistochemistry (IHC).
In this study, the patients were first divided into four groups by the MET expression level (0, 1+, 2+, and 3+) and then combined into two groups,
namely, MET negative (0, 1+) and MET positive (2+, 3+). See Figure \ref{fig:IHCPlot}.
The study concluded that the dual treatment was more efficacious in the MET positive patients but not in the MET negative patients, as compared to the single treatment.
The HR (between the dual treatment and the single treatment) was used as the efficacy measure when evaluating the treatment efficacy in each MET sub-population and the overall
population.
When combining the two IHC subgroups into a single group, a Cox model {\it ignoring the subgroup labels} (e.g., 2+ and 3+) was fitted and the HR from that model was used to measure
the efficacy for the \{2+, 3+\} combined group. Interestingly, Simpson's Paradox was observed, with the HR of the combined group (i.e., MET positive, a value of 0.53) being
bigger than the HR of each 2+ or 3+ subgroup (0.468 and 0.339, respectively).
The same approach was used when evaluating the treatment efficacy in the overall population when combining MET$^+$ and MET$^-$.

We now illustrate the two separate issues in measuring the treatment efficacy in this study,
which has been commonly observed from other similar clinical trials.

\subsection{Two separate issues} \label{subsec:issue}

The first issue is it turns out HR is {\it not} suitable
  for measuring efficacy in a mixture population because a mixture
  population typically will not have a constant HR, even if each of its
  constituent subgroup has a constant HR.

Abbreviate ``treatment'' and ``control'' with $Rx$ and $C$, respectively. Denote by $g^+$ and $g^-$ the two subgroups that may have differential efficacy.
Use $f(\cdot)$, $S(\cdot)$ and $\lambda(\cdot)$ to denote the density, survival and hazard functions. For example, $S^{Rx}_{g^+}(t)$ denotes the survival function of the $g^+$
patients who receive $Rx$ and $S^{Rx}(t)$ denotes the survival function of the entire patients who receive $Rx$.
By definition, the HR for the combined group is
\begin{equation}
\overline{HR}(t) = \frac{\lambda^{Rx}(t)}{\lambda^C(t)} = \frac{f^{Rx}(t)/S^{Rx}(t)}{f^C(t)/S^C(t)} = \frac{f^{Rx}(t)S^{C}(t)}{f^{C}(t)S^{Rx}(t)}, \label{HR_correct}
\end{equation}
with each density or survival function being a mixture of $g^+$ and $g^-$ corresponding density or survival functions. For example,
\begin{equation*}
f^{Rx}(t) = \gamma^+f^{Rx}_{g^+}(t)+(1-\gamma^+)f^{Rx}_{g^-}(t)
\end{equation*}
and
\begin{equation*}
S^{Rx}(t) = \gamma^+S^{Rx}_{g^+}(t)+(1-\gamma^+)S^{Rx}_{g^-}(t)
\end{equation*}
with $\gamma^+$ being the population prevalence of the $g^+$ subgroup (independent of the random assignment of patients to $Rx$ and $C$).
Even if each subgroup has a constant HR, i.e.,
$$HR_{g^-}=\frac{f^{Rx}_{g^-}(t)S^{C}_{g^-}(t)}{f^{C}_{g^-}(t)S^{Rx}_{g^-}(t)} \quad \mbox{and} \quad HR_{g^+} = \frac{f^{Rx}_{g^+}(t)S^{C}_{g^+}(t)}{f^{C}_{g^+}(t)S^{Rx}_{g^+}(t)}
$$
for all $t$, the HR for the combined group (equation \ref{HR_correct}) is {\em not} a constant in general, but rather a complex function of time $t$.
Thus, using constant HRs to represent the efficacy in subgroups and mixtures of subgroups is inappropriate.

A second issue is estimating efficacy in a combination of subgroups by ignoring subgroup labels.
For example, fitting a Cox PH model ignoring the IHC subgroup labels.
The resulting estimates are commonly referred to as {\em marginal means}.
We use a graph to illustrate how marginal means can be misleading.

\begin{figure}[tbp]
  \begin{center}
  \includegraphics[width=0.6\textwidth]{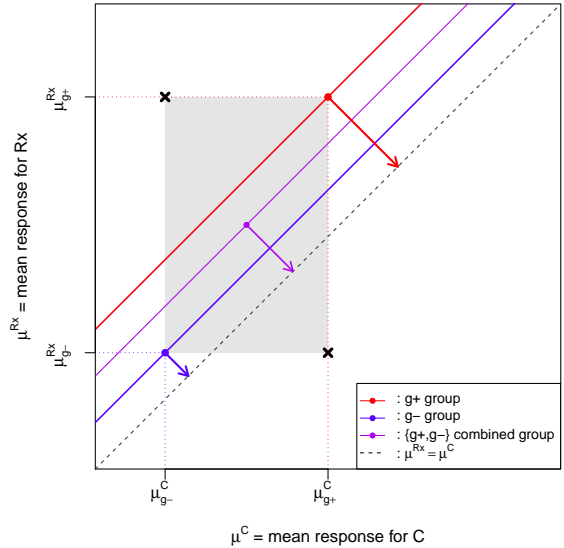}
  \caption{M\&M plot illustrating the Simpson's paradox.}
  \label{fig:ParadoxPlot}
  \end{center}
\end{figure}

Denote by $\mu^{Rx}$ and $\mu^{C}$ the true mean responses over the
  entire patient population {if the entire population had received treatment or control, respectively}.
Denote by $\mu_{g^+}^{Rx}$, $\mu_{g^-}^{Rx}$, $\mu_{g^+}^{C}$,
  $\mu_{g^-}^{C}$ the corresponding mean responses in the $g^+$ and $g^-$
  subgroups.
In what we call an M\&M plot (M\&M stands for `mean-and-mean' or `median-and-median'), the vertical axis represents $Rx$ value, while the horizontal axis represents $C$ value.
Suppose ($\mu_{g^+}^{Rx}$, $\mu_{g^+}^{C}$) and ($\mu_{g^-}^{Rx}$, $\mu_{g^-}^{C}$) are as drawn in Figure \ref{fig:ParadoxPlot}.
If efficacy is defined as the difference in mean response between $Rx$ and $C$, then efficacy in the $g^+$ and $g^-$ subgroups are perpendicular distances from those two points to the 45-degree line.
(Here and below it is understood that the distances are scaled by $\sqrt{2}$.)
If $g^+$ and $g^-$ are equally prevalent (50\% each), then true efficacy for the combined $g^+$ and $g^-$ population should be the perpendicular distance from the mid-point (of the
two dots) to the 45-degree line (denoted by the purple line and arrow).
However, in this particular finite sample, suppose $Rx$ patients are mostly $g^+$, while $C$ patients are mostly $g^-$, then the marginal means estimate for efficacy in the combined $g^+$ and $g^-$ population will
be close to the perpendicular distance from the upper left corner (denoted by `x' in the graph) of the shaded rectangle to the 45-degree line. Or similarly, if $Rx$ patients are
mostly $g^-$ and $C$ patients are mostly $g^+$, the the marginal means estimate for $g^+$ and $g^-$ combined will be close to the perpendicular distance from the lower right corner
`x' to the 45-degree line. Both indicate the Simpson's Paradox will occur and thus are illogical.

With these two issues present in assessing the treatment efficacy in a mixture population,
we thus propose a new principle, that of {\em subgroup mixable estimation},
to avoid inference that may be illogical when subgroups exist.

\section{Principle of Subgroup Mixable Estimation} \label{sec:SME}
Personalized medicine involves measuring efficacy in subgroups, and in
  mixtures of subgroups.
The Principle of Subgroup Mixable Estimation (SME) that we suggest is:

\begin{quote}
\it{Estimation should respect logical relationships among the parameters representing treatment efficacy in subgroups and their mixtures.}
\end{quote}


These logical relationships depend on the type of outcomes and the efficacy measures. They need to be established case by case and are not always ``generalizable''.
For example, when the outcome is continuous and the efficacy is measured as the difference of means (between $Rx$ and $C$), the relationship turns out to be an exact equality, where
the efficacy of the combined group is a weighted average of the subgroup's efficacy, with weights being the subgroups' prevalence. The corresponding estimation that reflects such a
logical relationship coincides with the LSmeans estimation. However, as LSmeans was originally developed for continuous outcomes, it should not be simply
``generalized'' to other types of outcomes when efficacy is defined differently. We have seen LSmeans being applied on the logarithm of HR from a Cox model to estimate the combined
group's HR, and on the logarithm of Relative Risk (RR) from a log linear model to estimate the combined group's RR. Unfortunately, none of them is correct since such these estimations
do not respect the logical relationships among the true efficacy parameters.

\subsection{Use of least squares means to adjust for design imbalance} \label{subsec:LSmeans}

The use of LSmeans for linear models dates back to at least
\cite{Goodnight1978}, who describe them as
\begin{quote}
\it{
Simply put, they are estimates of the class or subclass arithmetic
  means that would be expected had equal subclass numbers been
  obtainable.}
\end{quote}
Clearly, the use of LSmeans is to adjust for imbalance in the design when outcome is continuous,
not to directly estimate the efficacy of the combined subgroups.

Even though LSmeans can be thought of as least squares estimates of the true parameters in the model computed from $(X'X)^{-1}X'Y$,
  the following alternative description (given in \cite{Fleiss1986} for example) provides better insight.
It can best be understood in the setting of comparing the effect of $k$ doses of a
  drug with dose level $0$ effect, when there is also a block (i.e., other covariate) effect.
\begin{enumerate}
\item Within each block, estimate unbiasedly mean response at each dose (including dose 0) by its sample mean, noting the variance-covariance matrix of these estimates;
\item Within each block, estimate dose $i$ vs. dose $0$ effect, $i =
  1, \ldots, k,$ by taking differences of their respective sample means, deriving the variance-covariance matrix of these estimates from the variance-covariance matrix in step 1;
\item Calculate LSmeans of dose $i$ vs. dose $0$ effect combined across the blocks, $i =
  1, \ldots, k,$ by averaging the within block estimates, weighted by inverses of the variance-covariance matrices.
\end{enumerate}
Indeed, this is the adjustment for design imbalance behind the implementation of multiple comparisons within the LSmeans option of SAS Proc GLM and Proc Mixed, following \cite{Hsu1992} and \cite{Hsu1996}.

However, the LSmeans estimation has been ``broadly'' applied to directly compute the estimated treatment efficacy for the combined group (when subgroups exist) for different types of outcomes.
The thinking behind is so long as the ``efficacy'' for each subgroup has been estimated, by considering the subgroup as the ``block'', the efficacy for the overall group can be obtained via this weighted average approach. Such an approach misses two facts (1) the efficacy measure has to be suitable for a mixture population (while the ``natural parameter'' from the fitted model may not be a suitable efficacy measure) and (2) the estimation has to respect the logical relationships among parameters that represent the efficacy for subgroups and their mixtures.

\subsection{Importance of respecting logical relationships among parameters} \label{subsec:logic}

Before assessing efficacy between $Rx$ and $C$,
their treatment responses (e.g., change from baseline in HbA1c for an anti-diabetic drug) in each subgroup and their mixtures need to be estimated within $Rx$ and $C$ respectively.
Use the same notation as in the description of M\&M plot, we propose the general principle of SME.

\begin{pcp} \label{pcp:SME}
The three steps in the general principle of SME:
\begin{enumerate}
\item Within each treatment $Rx$ and $C$, estimate the response in each subgroup $\mu_{g^+}^{Rx}$, $\mu_{g^-}^{Rx}$ and
    $\mu_{g^+}^{C}$, $\mu_{g^-}^{C}$;
\item Within each treatment $Rx$ and $C$, estimate additionally $\mu^{Rx}$ and $\mu^{C}$, the responses in the mixture of the $g^+$ and $g^-$ subgroups;
\item Calculate efficacy ($Rx$ vs. $C$) in each subgroup, $g^+$ and $g^-$, and in their mixture, based on the pre-selected efficacy measure.
\end{enumerate}
\end{pcp}

Note that $\mu$ here denotes a general parameter for the treatment response, which may not necessarily be the mean. It can be median, response rate and etc. In Step 1,
adjusting for imbalance in sample sizes and other covariates (such as baseline measurements) can be done under a model for which the LSmeans technique suitably applies,
even if the model parameters involve transformations of  $\mu_{g^+}^{Rx}$, $\mu_{g^-}^{Rx}$, $\mu_{g^+}^{C}$, and $\mu_{g^-}^{C}$, so long as one can recover the estimates of these
original parameters that represent the treatment effect.
However, estimation of $\mu^{Rx}$ and $\mu^{C}$ in Step 2 needs to respect their logical relationships with $\mu_{g^+}^{Rx}$, $\mu_{g^-}^{Rx}$, and $\mu_{g^+}^{C}$, $\mu_{g^-}^{C}$.
While for continuous outcomes the models and parameter scales in Steps 1 and 2 can be the same,
for binary and time-to-event outcomes they are typically different, as we show below.

\subsubsection{Linear outcome with efficacy measured as a difference of means - a `special' case} \label{subsec:linear}

If efficacy is measured by the {\em difference} of expectation of treatment and
  control {outcomes}, and a higher mean response is better, then
\begin{equation*}
\mu_{g^{+}} = \mu_{g^+}^{Rx} - \mu_{g^+}^{C}, \quad \mu_{g^{-}} = \mu_{g^-}^{Rx} - \mu_{g^-}^{C}
\end{equation*}
represent efficacy of the treatment in the $g^+$ and
  $g^-$ subgroups.
If population prevalence of the $g^+$ subgroup is $\gamma^+$
  {(independent of the random assignment of patients to $Rx$ and
  $C$)}, then follow the principle of SME, the efficacy in the combined population is
\begin{equation}\label{eq.mean.ave}
\overline{\mu} = \mu^{Rx} - \mu^C =
[\gamma^+  \mu_{g^+}^{Rx}+ (1-\gamma^+) \mu_{g^-}^{Rx}] - [\gamma^+ \mu_{g^+}^{C}+ (1-\gamma^-)\mu_{g^-}^{C}]
= \gamma^+ \mu_{g^{+}}+(1-\gamma^+) \mu_{g^{-}}
\end{equation}
The last equation holds because the {weighted} average of the expected differences in
  the subgroups equals the expected difference in the mixture population.
Therefore, for continuous outcome modeled by a linear model with i.i.d. normal
errors, by the Gauss-Markov theorem, using linear combinations of LSmeans (of each subgroup's efficacy) respects the logical relationship among parameters and thus appropriately
estimate the efficacy in the combined population.

From the relationship (\ref{eq.mean.ave}),
one might get the impression that it is convenient to first estimate efficacy within each subgroup, and then estimate efficacy for the mixture population by taking average, weighted
by prevalence.
Some models can in fact be parameterized so that certain parameters in the model represent efficacy in the subgroups or their logarithms, making this possibility even more tempting.
To calculate efficacy for the mixture by averaging estimates of these parameters, weighted by prevalence, turns out to be hazardous if modeling involves a transformation (e.g., the
{\em log} transform), and/or if efficacy is defined as anything but a difference.
The case of binary outcome is a good illustration.


%
\subsubsection{Binary outcome with efficacy measured as a relative risk - a `general' case} \label{subsec:binary}

Let $p^{Rx}_{g^+(R)}$ represent the joint probability of a patient
  being assigned to treatment, belonging to the $g^+$ subgroup, and
  experiencing a positive response.  Let $p^{C}_{g^-(NR)}$ represent the
  corresponding joint probability for a patient being assigned to
  control, belonging to the $g^-$ subgroup, and {\em not} experiencing
  a positive response.
  The analogous joint probabilities, and their
  marginal sums, are displayed in Table \ref{table.response.prob}.
Note that appropriate mixing is on the response rate scale, within the $Rx$ and $C$ populations, using the fact that a mixture of Bernoulli distributions is a Bernoulli
distribution.
Table \ref{table.RR.example} gives a numerical illustration.

\begin{table}
\renewcommand\arraystretch{1.5}
\footnotesize{
\begin{center}
 \begin{tabular}{c|c|c|cp{5ex}|c|c|cp{5ex}|c|c|c}
  &\multicolumn{3}{c}{$g^+$ subpopulation}  & &\multicolumn{3}{c}{$g^-$ subpopulation} &&\multicolumn{3}{c}{population} \\
 \cline{2-4}\cline{6-8}\cline{10-12}
 & R & NR &&&R & NR & & &R &NR &\\
 \cline{1-4}\cline{6-8}\cline{10-12}
 Rx &$p^{Rx}_{g^+(R)}$&$p^{Rx}_{g^+(NR)}$ &$p_{g^+}^{Rx}$ & \centering + & $p^{Rx}_{g^-(R)}$&$p^{Rx}_{g^-(NR)}$ &$p_{g^-}^{Rx}$ & \centering = & $p^{Rx}_{(R)}$&$p^{Rx}_{(NR)}$
 &$p^{Rx}$\\
 \cline{1-4}\cline{6-8}\cline{10-12}
 C &$p^{C}_{g^+(R)}$&$p^{C}_{g^+(NR)}$ &$p_{g^+}^{C}$ & \centering + & $p^{C}_{g^-(R)}$&$p^{C}_{g^-(NR)}$ &$p_{g^-}^{C}$ & \centering = & $p^{C}_{(R)}$&$p^{C}_{(NR)}$ &$p^{C}$\\
 \cline{1-4}\cline{6-8}\cline{10-12}
 &$p_{g^+(R)}$&$p_{g^+(NR)}$ &$p_{g^+}$& &$p_{g^-(R)}$&$p_{g^-(NR)}$ &$p_{g^-}$& &$p_{(R)}$&$p_{(NR)}$& 1\\
 \end{tabular}
\caption{Probabilities of treatment assignment ($Rx$ or $C$),
  biomarker subgroup ($g^+$ or $g^-$), and response (responders (R) or
  non-responders (NR)).
The table on the right displays the correct probabilities when the $g^+$ and $g^-$ subgroups are combined, so that the sum of the probabilities in corresponding cells of the two
tables at the left equals the probability denoted in the corresponding cell of the right-hand table.}
  \label{table.response.prob}
\end{center}
}
\end{table}
%
%
\begin{table}
\renewcommand\arraystretch{1.5}
\footnotesize{
\begin{center}
 \begin{tabular}{c|c|c|cp{5ex}|c|c|cp{5ex}|c|c|c}
  &\multicolumn{3}{c}{$g_+$ subpopulation}  &
  &\multicolumn{3}{c}{$g_-$ subpopulation}
  &&\multicolumn{3}{c}{population} \\
 \cline{2-4}\cline{6-8}\cline{10-12}
 & R & NR &&&R & NR & & &R &NR &\\
 \cline{1-4}\cline{6-8}\cline{10-12}
 Rx & $8/86$ & $12/86$ & $20/86$
 & \centering + & $10/86$&$13/86$ &$23/86$
 & \centering = & $18/86$&$25/86$ &$43/86$\\
 \cline{1-4}\cline{6-8}\cline{10-12}
C &$3/86$&$17/86$ &$20/86$ &
\centering + & $12/86$&$11/86$ &$23/86$ &
\centering = & $15/86$&$28/86$ &$43/86$\\
 \cline{1-4}\cline{6-8}\cline{10-12}
 &$11/86$&$29/86$ &$40/86$& &$22/86$&$24/86$
 &$46/86$& &$33/86$&$53/86$& 1\\
 \end{tabular}
\caption{A numerical example illustrating probabilities of treatment assignment ($Rx$ or $C$), biomarker subgroup ($g^+$ or $g^-$), and response (responders (R) or
  non-responders (NR)).}
 \label{table.RR.example}
\end{center}
}
\end{table}

Relative Risk (RR) is a common measure of efficacy for binary outcomes.
Letting $RR_{g^+}$, $RR_{g^-}$ and $\overline{RR}$ denote RR for the $g^+$,
  $g^-$ subgroup and for overall population respectively, they are:
\begin{equation}
RR_{g^+} 
= \frac{ {^{p^{Rx}_{g^+(R)}}/_{p^{Rx}_{g^+}}}}{ ^{p^{C}_{g^+(R)}}/_{p^{C}_{g^+}}}
 = \frac{p^{Rx}_{g^+(R)} p^{C}_{g^+}}{ p^{C}_{g^+(R)} p^{Rx}_{g^+}}, \quad  RR_{g^-} = \frac{p^{Rx}_{g^-(R)} p^{C}_{g^-}}{ p^{C}_{g^-(R)} p^{Rx}_{g^-}}, \quad
\overline{RR} = \frac{p^{Rx}_{(R)} p^{C}}{ p^{C}_{(R)} p^{Rx}}.
\label{formula.RR}
\end{equation}
Note that in our case, the event associated with RR is a positive event (i.e., being a responder) instead of a negative event. Thus, the RR here refers to ``Relative Response''.

Binary outcome is commonly modeled by a log-linear model.
Applying LSmeans to this model will adjust for design imbalance in estimating model  parameters.
However parameterized, the complete set of parameters in the model are in 1-to-1 correspondence to response rates in the treatment $\times$ subgroup combinations.
Therefore, it is possible to follow the principle of SME, as follows.
\begin{pcp}
The three steps in the principle of SME for binary outcomes with RR as the efficacy measure.
\begin{enumerate}
\item First, estimate all the parameters in the log-linear model, using LSmeans to adjust for design imbalance, then transform to the response rate scale to get estimates of
    parameters in
    the two tables on the left of Table \ref{table.response.prob};
\item Calculate estimates of $p^{Rx}_{(R)} = p^{Rx}_{g^+(R)} + p^{Rx}_{g^-(R)}$ and $p^{C}_{(R)} = p^{C}_{g^+(R)} + p^{C}_{g^-(R)}$;
\item Calculate estimates of $RR_{g^+}$, $RR_{g^-}$, and $\overline{RR}$ based on (\ref{formula.RR}).
\end{enumerate}
\end{pcp}

We now show the hazard of not following the SME principle but mixing on the parameters from log-linear model directly, using the numbers in Table \ref{table.RR.example}
for illustration.
Mixing logarithms of RR by prevalence results in
$$\frac{20}{43} \times log \left( \frac{8}{3} \right)+
\frac{23}{43} \times log \left( \frac{10}{12} \right) \ne
log \left( \frac{18}{15} \right)
$$
while mixing RR by prevalence results in
$$\frac{20}{43} \times \left( \frac{8}{3} \right)+
\frac{23}{43} \times \left( \frac{10}{12} \right) =
\frac{145}{86} \ne \frac{18}{15}.
$$

To provide insight, we note the true relative risk $\overline{RR}$ is calculated as
\begin{equation}
\frac{p^{C}_{g^-(R)}}{p^{C}_{g^-(R)}+p^{C}_{g^+(R)}} \times
\frac{p^{Rx}_{g^-(R)} p^{C}_{g^+(R)}}{p^{C}_{g^-(R)} p^{C}_{g^+(R)}} +
\frac{p^{C}_{g^+(R)}}{p^{C}_{g^-(R)}+p^{C}_{g^+(R)}} \times
\frac{p^{Rx}_{g^+(R)} p^{C}_{g^-(R)}}{p^{C}_{g^+(R)} p^{C}_{g^-(R)}}
= \frac{p^{Rx}_{g^-(R)}+p^{Rx}_{g^+(R)}}{p^{C}_{g^-(R)}+p^{C}_{g^+(R)}},
\label{eq.RRmix}
\end{equation}
so $\overline{RR}$ is not a mixture of $RR_{g^+}$ and $RR_{g^-}$ weighted by population prevalence of the subgroups, nor it is in the logarithm scale.
It is a mixture of $RR_{g^+}$ and $RR_{g^-}$ weighted by population proportion of control responders who are $g^+$ and $g^-$ respectively (see \cite{Tang2013}).
The correct estimation should always follow the subgroup mixable principle, which respects the logical relationship among the efficacy parameters.

\section{Subgroup Mixable Estimation for Time-to-event Outcomes} \label{sec:TTE}

We now concentrate on the time-to-event outcome setting. Multiple papers offered different ways in assessing the
combined group's efficacy using HR under a Cox PH model, where the model contains treatment, marker and treatment-by-marker interaction effects (and possible additional covariates).
For example, in the thresholding paper of \cite{Jiang2007} and
in the motivating example we have introduced \citep{Spigel2013},
  efficacy in the combined group is estimated by marginal
  means.
While in \cite{Kalser2011}, it is claimed that the log of HR for the
  combined group $\log(\overline{HR})$ can be approximated by the
  following formula:
\begin{equation}
\log(\overline{HR}) = f^+ \log(HR_{g^+}) + (1-f^+) \log(HR_{g^-}), \label{HR_Genetech}
\end{equation}
where $f^+=d_{+}/d$ is the proportion of total events that occur in $g^+$ group. On the other hand, SAS procedure Proc PHREG (for fitting Cox PH models) \citep{SAS} provides a LSmeans estimate for the HR of the combined group (with LSMEANS statement), which in our case is
\begin{equation}
\log(\overline{HR}) = \log(HR^{LS,Rx})-\log(HR^{LS,C}). \label{HR_LSMEANS}
\end{equation}
Here $\log(HR^{LS,Rx})$ and $\log(HR^{LS,C})$ denote the log of LSmeans estimates of HR in the treatment and control group (as compared to the reference group, which is the $g^-$ group receiving $C$), respectively.
Clearly, none of them is correct as they all yield a constant HR. Since HR is not suitable for a mixture population, we suggest suitable efficacy measures together with an estimation procedure that follows the SME principle.

\subsection{Efficacy measures that are suitable for mixture population} \label{subsec:eff_TTE}

Median or mean survival times are often of interest in oncology trials with time-to-event outcomes.
The difference or the ratio of the median (or mean) survival times between $Rx$ and $C$ provides direct information on the relative treatment effects.
For example, if the median survival time for patients randomized to $Rx$ is 18 months and the median survival time for patients randomized to $C$ is 12 months.
Then $Rx$ extends the median survival time for 6 (=18-12) months as compared to $C$, or 1.5 times (=18/12) of $C$. We now show that, under a Weibull model, both the difference of median (or mean) survival and the ratio of median (or mean) survival are suitable for measuring efficacy in subgroups and their mixtures. In particular, in the parameter space of these efficacy measures, it is always guaranteed that the Simpson's paradox cannot occur, that is, the efficacy of the mixture always stays within the interval of the subgroups' efficacy.

\begin{thm} \label{thm:weibull}
Assume the time-to-event data fit the following Cox PH model:
\begin{equation} \label{model:Weibull}
h(t|Trt,M)= h_0(t)\exp\{\beta_1 Trt + \beta_2 M + \beta_3 Trt \times M \},
\end{equation}
where $Trt=0$ ($C$) or $Trt=1$ ($Rx$), $M=0$ ($g^-$) or $M=1$ ($g^+$), and $h_0(t)=h(t|C,g^-)$ is the hazard function for the $g^-$ subgroup receiving $C$.
Further assume that the survival function $S_0(t)$ for $C,g^-$ is from a Weibull distribution with scale $\lambda$ and shape $k$, i.e.,
$$ S_0(t) (= S_{g^-}^C(t)) = e^{-(t/\lambda)^k}, \quad t \ge 0.$$
If efficacy is defined as the difference of median (or mean) survival (between $Rx$ and $C$), or by the ratio of median (or mean) survival (between $Rx$ and $C$), then
the efficacy of $g^-$, $g^+$, and their mixture can all be represented by a function of the five model parameters $(\lambda, k, \beta_1, \beta_2, \beta_3)$. More importantly, the
efficacy of the mixture is always guaranteed to stay within the interval of the subgroups' efficacy.
\end{thm}

{Remark}. The Weibull model above (\ref{model:Weibull}) does not include other covariates for adjustment. In practice, the model can include additional covariates that are known to be associated with the outcome (e.g., baseline characteristics). The result of Theorem \ref{thm:weibull} still holds with these additional model parameters. In below, we provide the proof for the case without additional covariates.

\begin{proof}
We use ratio of median survival as an example. The other efficacy measures (i.e., ratio of mean or difference of median or mean) follow from a similar argument. Denote by $\nu^{Rx}$ and $\nu^{C}$ the true median
survival times over the entire patient population (randomized to $Rx$ and $C$ respectively). Denote by $\nu_{g^+}^{Rx}$, $\nu_{g^-}^{Rx}$, $\nu_{g^+}^{C}$, $\nu_{g^-}^{C}$ the
corresponding median survival times in the $g^+$ and $g^-$ subgroups. Denote $\theta_1=e^{\beta_1}$, $\theta_2=e^{\beta_2}$ and $\theta_3 = e^{\beta_3}$. Note that $\theta_1,
\theta_2, \theta_3$ all $>0$.

By the PH property, the survival function for each of the subgroups has the following form
\begin{eqnarray*}
&& S_{g^-}^{C}(t) = e^{-(t/\lambda)^k}, \quad S_{g^-}^{Rx}(t) = e^{-\theta_1(t/\lambda)^k}, \\
&& S_{g^+}^{C}(t) = e^{-\theta_2(t/\lambda)^k}, \quad S_{g^+}^{Rx}(t) = e^{-\theta_1\theta_2\theta_3(t/\lambda)^k}.
\end{eqnarray*}
Straightforward calculation gives the median survival for each subgroup as follows
\begin{equation}
\nu_{g^+}^{Rx} = \lambda (\frac{\log2}{\theta_1\theta_2\theta_3})^{1/k}, \quad \nu_{g^+}^{C} = \lambda (\frac{\log2}{\theta_2})^{1/k},
\quad \nu_{g^-}^{Rx} = \lambda (\frac{\log2}{\theta_1})^{1/k}, \quad \nu_{g^-}^{C} = \lambda (\log2)^{1/k}. \label{eq:median_subgroup}
\end{equation}
Then the ratios of median for $g^+$ and $g^-$ are
\begin{equation}
r_{g^+} = (\theta_1\theta_3)^{-1/k}  \quad \mbox{and} \quad r_{g^-} = \theta_1^{-1/k},
\label{eq:ratio_subgroup}
\end{equation}
which are functions of $(k,\theta_1,\theta_3)$.
For the mixture of $g^+$ and $g^-$, the median survival times for $Rx$ and $C$ are the solutions for the following two equations respectively.
\begin{eqnarray}
&& t=\nu^{Rx}: \ (1-\gamma^+) e^{-\theta_1(t/\lambda)^k}+ \gamma^+ e^{-\theta_1\theta_2\theta_3(t/\lambda)^k} = 0.5,
\label{eq:mix_Rx} \\
&& t=\nu^{C}: \ (1-\gamma^+) e^{-(t/\lambda)^k}+ \gamma^+ e^{-\theta_2(t/\lambda)^k} = 0.5. \label{eq:mix_C}
\end{eqnarray}
Then the ratio of median for the mixture group $\overline{r} \equiv \nu^{Rx}/\nu^{C}$ is an implicit function of $(\lambda,k,\theta_1,\theta_2,\theta_3)$.

Now, we show that $\overline{r}$ is between $r_{g^-}$ and $r_{g^+}$. Let $t=\nu^C r_{g^-}=\nu^C \theta_1^{-1/k}$ and plug into the left side of equation (\ref{eq:mix_Rx}), we have
\begin{eqnarray}
&& (1-\gamma^+) e^{-\theta_1(\nu^C \theta_1^{-1/k} /\lambda)^k}+ \gamma^+ e^{-\theta_1\theta_2\theta_3(\nu^C \theta_1^{-1/k}/\lambda)^k} \label{eq:mix_proof1} \\
&& \quad = (1-\gamma^+) e^{-(\nu^C/\lambda)^k}+\gamma^+ e^{-\theta_2\theta_3(\nu^C/\lambda)^k} \label{eq:mix_proof2}.
\end{eqnarray}
The first term in equation (\ref{eq:mix_proof2}) equals the first term on the left side of (\ref{eq:mix_C}) with $\nu^C$ plugged in. Therefore, whether $(\ref{eq:mix_proof2}) > 0.5$ or
$<0.5$ depends on whether $\theta_3 < 1$ or $ > 1$. Without loss of generosity, assume $\theta_3>1$. Then by the property that the all survival functions are non-increasing
functions, comparing (\ref{eq:mix_Rx}) with (\ref{eq:mix_proof1}), we have
$$  \nu^{Rx} > \nu^C \theta_1^{-1/k} = \nu^C r_{g^-}. $$
Thus, $\overline{r} = \nu^{Rx}/\nu^C > r_{g^-}$. With a similar argument, we can show that $\overline{r} < r_{g^+}$ (if $\theta_3>1)$. Hence, we have shown that the ratio of median
survival time for the mixture population is within the interval of the ratios for the subgroups and each ratio can be represented by a function of
$(\lambda,k,\theta_1,\theta_2,\theta_3)$ either explicitly or implicitly.
\end{proof}

With suitable efficacy measures for the time-to-event outcomes chosen, we now provide the corresponding principle of SME and illustrate the estimation results using the graphical tool we develop, namely the M\&M plot.

\subsection{Subgroup mixable estimation with the use of M\&M plot} \label{subsec:MMplot}
We use ratio of median survival as the efficacy measure, with the Weibull model, to illustrate the key steps of SME.
\begin{pcp}
The three steps in the principle of SME for time-to-event outcomes, with ratio of median survival as the efficacy measure under a Weibull model are as follows.
\begin{enumerate}
\item First, estimate all the parameters in the Weibull model (e.g., $(\lambda,k,\theta_1,\theta_2,\theta_3)$ in the case without additional covariates).
\item Then, within each treatment $Rx$ and $C$, compute the median survival estimates for $g^+$ and $g^-$ and their mixture based on equations (\ref{eq:median_subgroup}), (\ref{eq:mix_Rx}) and (\ref{eq:mix_C}), and compute their estimated variance covariance matrix by the Delta method.
\item Finally, calculate the ratio estimates (between $Rx$ and $C$) for $g^+$ and $g^-$ and their mixture, and compute their estimated variance covariance matrix based on the Delta method.
\end{enumerate}
\end{pcp}

In Step 2, the Delta method for implicitly defined random variables \citep{Benichou1989} needs to be applied since the median survival for the combined group in $Rx$ and $C$ are implicitly defined by equations (\ref{eq:mix_Rx}) and (\ref{eq:mix_C}). In Step 3, the asymptotic normal approximation in the Delta method can be applied on the logarithm of ratios (instead of the original ratios). Then transform back to the original scale when computing the simultaneous confidence intervals for the ratios. In this way, the confidence intervals are guaranteed to be positive.

\begin{figure}[tbp] 
  \begin{center}
  \includegraphics[width=1\textwidth]{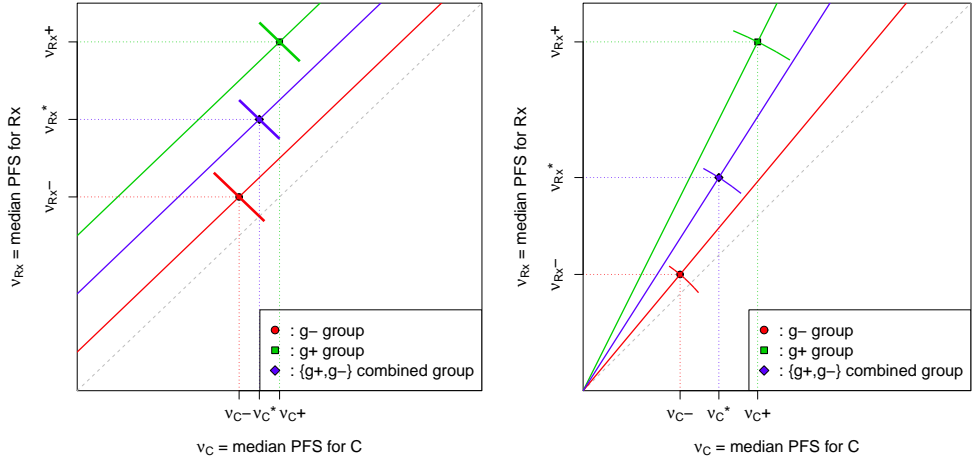}
  \caption{An example of M\&M plot when difference or ratio of median PFS is the efficacy measure.}
  \label{fig:RatDifPlot}
  \end{center}
\end{figure}

We now illustrate how to use the M\&M plot to display a SME result for two efficacy measures, difference and ratio of the median survival, respectively. The left panel of Figure \ref{fig:RatDifPlot} is an example M\&M plot when difference of median PFS is chosen as the efficacy measure. The x-axis represents the median PFS for $C$ and the y-axis represents the median PFS for $Rx$. The estimated median PFS within $C$ and $Rx$ for $g^-$, $g^+$ and the combined group are denoted by the red circle, green square and blue diamond, respectively. The intercept of the corresponding colored line gives the estimated efficacy value for each group (e.g., difference $\nu_{Rx^-}-\nu_{C^-}$ for $g^-$ group). The three line segments that are symmetric and perpendicular to the 45-degree line represent the simultaneous 95\% confidence intervals for the difference of medians. A segment that does not cross the diagonal line indicates a significant efficacy. In this example, all three groups have positive efficacy estimates (differences $>0$) with significance in the $g^+$ and the combined groups. The right panel of Figure \ref{fig:RatDifPlot} is an example M\&M plot when ratio of median PFS is chosen as the efficacy measure. Different from the left panel figure, the slope of each line that passes through the origin gives the estimated efficacy value for each group (e.g., ratio $\nu_{Rx^-}/\nu_{C^-}$ for $g^-$ group). The three arcs around the dot (or square/diamond) represent the simultaneous 95\% confidence intervals for the ratio of medians. It is interesting to point out that, different from the line segment in the left panel figure, the length of the arc is not symmetric around the dot (or square/diamond), however the degree of the arc is symmetric around the line. An arc that does not cross the diagonal line indicates a significant efficacy. In this example, all three groups have positive efficacy estimates (ratios $>1$) with significance in the $g^+$ and the combined groups.

\subsection{Simulation studies} \label{subsec:simu}
We conduct simulations to investigate the finite sample performance of the proposed Weibull-model-based SME under different scenarios. Survival times are generated from model (\ref{model:Weibull}) with the Weibull distribution scale $\lambda=50$ and shape $k=1.25$. The scale parameter is chosen so that the unit of simulated survival times is in weeks. Three different sets of $(\beta_1,\beta_2,\beta_3)$ values are considered, which are (a) $(\beta_1,\beta_2,\beta_3) = (0.5, -1,-1) $; (b) $(\beta_1,\beta_2,\beta_3) = (-0.1, -0.5,-0.5) $; and (c) $(\beta_1,\beta_2,\beta_3) = (-0.5, -1,0) $. The true median survival and their ratios or differences in $g^-$, $g^+$ and $\{g^-,g^+\}$ combined are listed in Table \ref{table.simu.para} for each scenario. The hazard ratios (between $Rx$ and $C$) for each subgroup are also listed. In scenario (a), treatment $Rx$ is not efficacious in $g^-$ group (indicated by shorter median survival as compared to $C$), while it is efficacious in $g^+$ group (indicated by longer median survival as compared to $C$). In scenario (b), $Rx$ is efficacious in both $g^-$ and $g^+$ groups but the HRs are different. Finally in scenario (c), $Rx$ is efficacious in both groups with the same HRs. In all three scenarios, the subjects in $g^-$ group have shorter mean survival than the $g^+$ group regardless of the treatment, indicated by the negative $\beta_2$ value. The censoring times are generated from a uniform distribution $U[a,b]$ with $a$ and $b$ chosen to yield two censoring rates 20\% and 50\%. Two different prevalences of marker groups are considered: $g^- : g^+ = 1:1$ and $g^- : g^+ = 1:3$. Equal randomization between $Rx$ and $C$ is assumed. We simulate 1000 runs for each setting and report the simulation results for a total sample size of 400 in Table \ref{table.simu.diff} (for difference of medians) and Table \ref{table.simu.ratio} (for ratio of medians). For the ratio, we apply the Delta method (for implicitly defined random variables) on two different scales: original ratio scale and log(ratio) scale when computing the confidence intervals. The reported confidence intervals are simultaneously 95\% based on asymptotic multivariate normal distributions.

When difference is the efficacy measure, we can see from Table \ref{table.simu.diff} that the bias of the difference estimates for each subgroup or their mixture is minimal for all three scenarios, even with a high censoring rate. The coverage probability is around 95\% for all scenarios. The confidence intervals of the efficacy estimates reflect the correct inference for all three groups. It is worth to point out that if the two individual marker groups demonstrate opposite efficacy, additional caution is required to conclude which group(s) should the treatment target for (e.g., scenario (a)). In such a situation, the combined group's efficacy estimate may be dominated by the subgroup that is more prevalent. So a positive efficacy in the combined group may not indicate the treatment is efficacious for the entire population. That's why all three groups' efficacy need to be evaluated carefully in order for a correct decision on drug development. For scenario (a), the result supports to target $g^+$ group only. For scenario (b), the result supports to target $\{g^-, g^+\}$ combined but realizing that only the $g^+$ group shows a significantly positive efficacy. Finally for scenario (c), the result also supports to target $\{g^- g^+\}$ combined with both $g^-$ and $g^+$ showing significantly positive efficacy.

When ratio is the efficacy measure, Table \ref{table.simu.ratio} shows that the bias of the ratio estimates is also minimal for each subgroup or their mixture across all the scenarios. The coverage probabilities under both scales are close to 95\% for all scenarios, with the ones under the log scale showing a slightly better accuracy. For each scenario, the confidence intervals of the ratio estimates support the same efficacy conclusion as the difference result. Overall, the above simulation studies demonstrate that the proposed SME performs well with both efficacy measures when sample size is moderate.

\begin{table}
\caption{True efficacy values in three simulation scenarios. The true median survival times in each treatment and marker group are provided. Their ratio and difference between $Rx$ and $C$ for each subgroup and the mixture are also presented, as well as the hazard ratio for each subgroup.}
\begin{center}
 \begin{tabular}{cccccc}
  \cline{1-6}
Population & $C$ & $Rx$ & $Rx/C$ & $Rx-C$ & HR \\
 \cline{1-6}
{Scenario (a)} \\
 $g^-$ & 37.3 & 25.0 & 0.67 & -12.3 & 1.65 \\
 $g^+$ & 83.0 & 123.8 & 1.5 & 40.8 & 0.61 \\
 $\{g^-, g^+\}$ & 54.0 & 49.4 & 0.9 & -4.6 & -- \\
{Scenario (b)} \\
  $g^-$ & 37.3 & 40.0 & 1.1 & 2.7 & 0.90 \\
 $g^+$ & 55.6 & 89.9 & 1.6 & 34.3 & 0.55 \\
 $\{g^-, g^+\}$ & 45.2 & 58.5 & 1.3 & 13.3 & -- \\
{Scenario (c)} \\
  $g^-$ & 37.3 & 55.6 & 1.5 & 18.3 & 0.61 \\
 $g^+$ & 83.0 & 123.8 & 1.5 & 40.8 & 0.61 \\
 $\{g^-, g^+\}$ & 54.0 & 80.5 & 1.5 & 26.5 & -- \\
 \cline{1-6}
 \end{tabular}
 \label{table.simu.para}
\end{center}
\end{table}

\begin{table}
\caption {Summary of the simulation statistics with efficacy being measured as the difference of median survival times.
The average bias of the efficacy estimate is reported for each group with average 95\% simultaneous CIs (sCIs) provided in parentheses. The coverage probabilities (CP) are also presented. }
\begin{center}
\begin{tabular}{lcccc}
\hline
$n_g^- : n_g^+$  & Difference $g^-$ & Difference $g^+$ & Difference $\{g^-,g^+\}$ & CP \\
(Censor Rate) & Bias (95\% sCI) & Bias (95\% sCI) & Bias (95\% sCI) &  \\
\hline
{Scenario (a)} \\
200:200 (20\%) & -0.096 (-20.85,-3.93) & 0.487 (4.13,78.49) & -0.091 (-15.42,6.07) & 0.952\\
200:200 (50\%) & -0.073 (-22.64,-2.09) & 1.549 (-13.26,98.00) & 0.028 (-18.50,9.39) & 0.960\\
100:300 (20\%) & -0.052 (-23.99,-0.71) & 0.161 (13.93,68.03) & 0.435 (0.51,34.30) & 0.947\\
100:300 (50\%) & -0.321 (-25.77, 0.54) & 0.733 (3.53,79.58) & 0.683 (-5.52,40.82) & 0.962\\
{Scenario (b)} \\
200:200(20\%) & -0.336 (-7.96,13.50) & 0.359 (9.54,59.73) & -0.214 (2.04,24.04) & 0.962\\
200:200 (50\%) & -0.258 (-10.45,16.14) & 1.483 (0.76,70.76) & -0.002 (-1.02,27.53) & 0.962\\
100:300 (20\%) & 0.108 (-11.66,18.08) & 0.738 (15.70,54.33) & 0.519 (9.75,36.39) & 0.953\\
100:300 (50\%) & -0.046 (-14.82,20.94) & 0.689 (9.05,60.88) & 0.362 (5.50,40.33) & 0.953\\
{Scenario (c)} \\
200:200 (20\%) & -0.056 (5.35,31.23) & -0.421 (4.94,75.86) & -0.269 (11.82,40.72) & 0.941\\
200:200 (50\%) & 0.451 (2.48,35.10) & 1.342 (-8.23,92.55) & 0.483 (7.81,46.24) & 0.963\\
100:300 (20\%) & -0.303 (0.24,35.84) & 0.610 (14.35,68.51) & 0.166 (15.04,51.27) & 0.947\\
100:300 (50\%) & 0.302 (-2.49,39.78) & -0.364 (3.78,77.13) & -0.178 (8.86,56.76) & 0.958\\
\hline
\end{tabular}
\label{table.simu.diff}
\end{center}
\end{table}

\newpage
\begin{landscape}
\begin{table}
\caption {Summary of the simulation statistics with efficacy being measured as the ratio of median survival times.
The average bias of the efficacy estimate is reported for each group with average 95\% simultaneous CIs (sCIs) provided in parentheses. The coverage probabilities (CP) are also presented.
The CIs and CP are reported separately for two different Delta methods, one based on original ratio scale and the other based on log(ratio) scale.
The bias of the efficacy estimate is the same for both scales and thus omitted in the log scale result.}
\begin{center}
\begin{tabular}{lccccccccc}
\hline
$n_g^- : n_g^+$  & \multicolumn{4}{c}{original scale}  & & \multicolumn{4}{c}{log scale} \\
\cline{2-5} \cline{7-10}
(Censor Rate) & $g^-$  & $g^+$ & $\{g^-,g^+\}$ ) & CP & & $g^-$  & $g^+$  & $\{g^-,g^+\}$  & CP \\
\hline
{Scenario (a)} \\
200:200 (20\%) & 0.003 (0.50,0.85) & 0.018 (1.00,2.02) & 0.002 (0.73,1.11) & 0.933 & & (0.52,0.88) & (1.08,2.12) & (0.75,1.13) & 0.945\\
200:200 (50\%) & 0.006 (0.46,0.89) & 0.040 (0.78,2.28) & 0.006 (0.68,1.17) & 0.949 & & (0.49,0.93) & (0.94,2.50) & (0.71,1.20) & 0.957\\
100:300 (20\%) & 0.008 (0.43,0.93) & 0.007 (1.12,1.87) & 0.010 (0.99,1.54) & 0.931 & & (0.47,0.98) & (1.17,1.93) & (1.02,1.57) & 0.940\\
100:300 (50\%) & 0.005 (0.40,0.95) & 0.020 (0.99,2.03) & 0.017 (0.90,1.64) & 0.938 & & (0.45,1.01) & (1.07,2.13) & (0.95,1.70) & 0.951\\
{Scenario (b)} \\
200:200 (20\%) & -0.002 (0.79,1.38) & 0.015 (1.11,2.16) & -0.001 (1.02,1.56) & 0.950 & & (0.82,1.42) & (1.18,2.25) & (1.05,1.59) & 0.955\\
200:200 (50\%) & 0.005 (0.72,1.46) & 0.043 (0.94,2.38) & 0.007 (0.95,1.65) & 0.942 & & (0.78,1.53) & (1.08,2.56) & (1.00,1.70) & 0.956\\
100:300 (20\%) & 0.018 (0.69,1.52) & 0.022 (1.23,2.05) & 0.016 (1.16,1.77) & 0.935 & & (0.76,1.61) & (1.28,2.11) & (1.19,1.81) & 0.946\\
100:300 (50\%) & 0.019 (0.60,1.60) & 0.021 (1.10,2.17) & 0.014 (1.07,1.86) & 0.940 & & (0.70,1.74) & (1.18,2.27) & (1.12,1.92) & 0.944\\
{Scenario (c)} \\
200:200 (20\%) & 0.005 (1.09,1.90) & 0.005 (1.01,1.99) & -0.001 (1.18,1.80) & 0.922 & & (1.14,1.96) & (1.08,2.07) & (1.21,1.84) & 0.937\\
200:200 (50\%) & 0.023 (1.01,2.02) & 0.032 (0.84,2.21) & 0.015 (1.10,1.91) & 0.952 & & (1.09,2.12) & (0.97,2.39) & (1.15,1.98) & 0.961\\
100:300 (20\%) & 0.010 (0.94,2.06) & 0.014 (1.13,1.88) & 0.007 (1.19,1.81) & 0.928 & & (1.03,2.18) & (1.17,1.94) & (1.22,1.85) & 0.948\\
100:300 (50\%) & 0.028 (0.86,2.18) & 0.004 (1.00,1.99) & 0.002 (1.09,1.90) & 0.938 & & (0.99,2.34) & (1.07,2.09) & (1.14,1.96) & 0.960\\
\hline
\end{tabular}
\label{table.simu.ratio}
\end{center}
\end{table}
\end{landscape}

\section{Application to the Motivating Example} \label{sec:app}

We ``reverse engineer'' the data from the motivating example \citep{Spigel2013} and apply the proposed SME method on it. Specifically, the PFS data for each MET$^+$ or MET$^-$ group are
generated based on the number of events, median survival, hazard ratio, the number of patients at risk for every three months and the Kaplan-Meier survival curves  provided in Figure
2(B) and 2(C) of \cite{Spigel2013}. MET$^+$ and MET$^-$ patients are further separated into IHC 2+ \& 3+ and IHC 0 \& 1+ respectively based on the analogous survival information from Figure A2(A) of the paper.

\begin{figure}
\begin{center}
\includegraphics[width=1\textwidth]{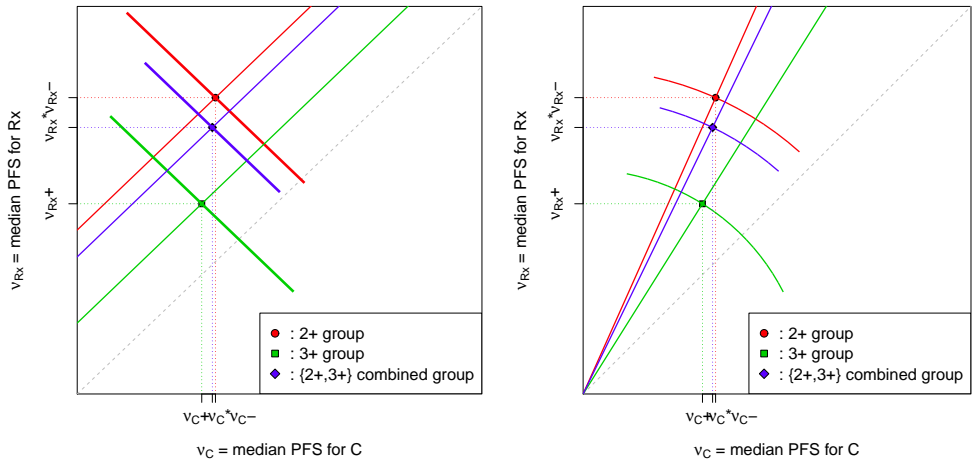}
\caption{The M\&M plots demonstrating the SME results on ``reverse engineered'' PFS data for ICH 2+ and 3+ patients in \cite{Spigel2013}.}
\end{center}
  \label{fig:RatDifPlot.JCO}
\end{figure}

We apply the SME method on the ``reversed engineered'' MET$^+$ data and compute the efficacy estimates based on difference of median survival and ratio of median survival for each subgroup and their mixture. The results are provided in Table \ref{table.ratdif.ex}. We also display the estimation results on the M\&M plot for each efficacy measure in Figure \ref{fig:RatDifPlot.JCO}. Both IHC 2+ and 3+ groups show positive efficacy (indicated by positive difference or $>1$ ratio in median survival). The combined group IHC $\{2+,3+\}$ also shows positive efficacy. However, only the combined group is statistically significant when difference is used as the efficacy measure (the 2+ group is marginally significant as it just crosses the 45-degree diagonal line)  while both the 2+ and the combined group are statistically significant when ratio is used as the efficacy measure. Note that with both efficacy measures, the estimated efficacy for the combined group stays between the estimated efficacy for the individual subgroup, which is clearly shown by the position of the lines in the M\&M plots. It is also worthwhile to point out that, with the proposed method, the 2+ group shows a better efficacy than the 3+ group (indicated by both the difference and the ratio); while with the approach in \cite{Spigel2013}, the 3+ group has a smaller estimated HR than the 2+ group (which tends to indicate the 3+ group receives more efficacy than the 2+ group).

\begin{table}
\caption{The SME result on ``reverse engineered'' PFS data of MET$^+$ patients in \cite{Spigel2013}. Median survival estimates for each subgroup and their mixture are provided. Difference and ratio of median survival estimates are also provided, together with the 95\% simultaneous CIs.}
\begin{center}
 \begin{tabular}{ccccc}
  \cline{1-5}
 Population & Median $C$ & Median $Rx$ & Difference and 95\% sCI & Ratio and 95\% sCI \\
 \cline{1-5}
 $2+$ & 2.22 (n=25) & 4.97 (n=26) & 2.75 (-0.11, 5.60) & 2.24 (1.12, 4.48) \\
 $3+$ & 2.00 (n=6)&  3.19 (n=9) & 1.19 (-1.76, 4.13) & 1.59 (0.51, 5.05) \\
 $\{2^+, 3^+\}$ & 2.17 (n=31) & 4.47 (n=35) & 2.30 (0.13, 4.47) & 2.06 (1.14, 3.74) \\
 \cline{1-5}
 \end{tabular}
 \label{table.ratdif.ex}
\end{center}
\end{table}

\section{Discussions} \label{sec:diss}

\subsection{Prognostic or predictive?}

The definition of ``prognostic'' or ``predictive'' biomarkers can be easily found in the literature. The prognostic biomarker is a disease-related biomarker, and it provides information on how such a disease may develop or progress in a patient population regardless of the type of treatment. While the predictive biomarker is a drug-related biomarker, it helps to assess whether a particular treatment will be more effective in a specific patient population. In personalized medicine, predictive biomarkers are the ones of interest.
However, the definitions usually do not specify how the effectiveness or equivalently, the efficacy of a treatment is measured. In fact, whether a biomarker is prognostic or predictive depends on the efficacy measure. We use the following example to illustrate.

\begin{figure}
\begin{center}
\includegraphics[width=0.6\textwidth]{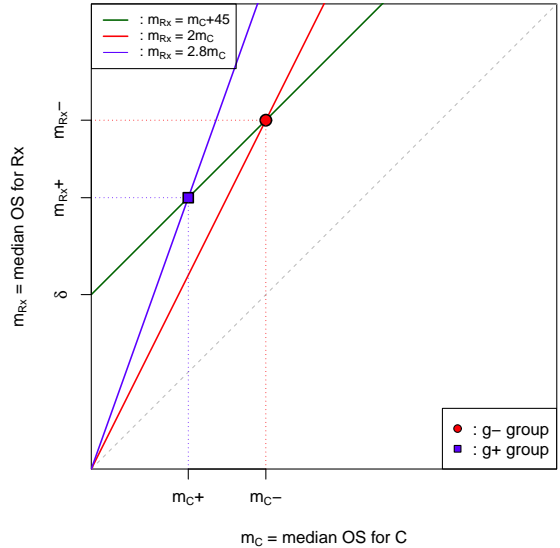}
\caption{The M\&M plot illustrating whether a biomarker is prognostic or predictive depends on the efficacy measure.}
\label{fig:DifRat.PredProg}
\end{center}
\end{figure}

Assume a biomarker divides the patient population into two groups $g^-$ and $g^+$ and each of them receive $Rx$ or $C$ randomly. The median OS for the $g^-$ group is 45 weeks if receiving $C$ and 90 weeks if receiving $Rx$. While the median OS for the $g^+$ group is 25 weeks if receiving $C$ and 70 weeks if receiving $Rx$. Is the marker prognostic or predictive? The answer will be different depending how the efficacy is measured. We plot the data on the M\&M plot and display in Figure \ref{fig:DifRat.PredProg}. If the efficacy is measured by the difference in median OS, both marker groups demonstrate the same treatment efficacy ($Rx-C=45$ weeks), indicated by the green 45-degree line in the figure. Therefore, the marker is not predictive for this particular treatment $Rx$. It is prognostic instead since the $g^+$ patients have a worse median survival outcome as compared to the $g^-$ patients, regardless whether they receive $C$ or $Rx$. However, if the efficacy is measured by the ratio in median OS, the $g^+$ group demonstrates a better efficacy than the $g^-$ group ($Rx/C=2.8$ in $g^+$ vs $Rx/C=2$ in $g^-$). Therefore, the marker is predictive for $Rx$ in this case.

\subsection{Extension and application of the subgroup mixable estimation}
With moderate effort, the current SME principle and its corresponding estimation procedure introduced in this work can be extended to handle $m$ subgroups with $m>2$. The user can then specify which subgroups are of interest to be combined. Additional covariates such as baseline characteristics can be also included in the model to adjust for. These extensions are under our current development.

The proposed SME principle has a broad application. As a closing remark, we present two important applications where inference on treatment efficacy in both subgroups and their mixtures are necessary in discovering personalized medicine.

Application 1: Pharmacogenomics studies that use genes or SNPs (Single Nucleotide Polymorphisms) known to be in the biological pathway of the action of that drug to identify targeted patient population.  The decision to make depends on the type of genetic effect and its effect size. For example, for a SNP with variants denoted by $AA$, $Aa$, and $aa$, suppose the $a$ allele is beneficial and it is dominant. Then the difference in treatment efficacy between the combined \{$Aa$, $aa$\} subgroup and $AA$ subgroup is its dominant effect size. Similarly, if $a$ is recessive, then the difference in efficacy between $aa$ subgroup and the combined \{$AA$, $Aa$\} subgroup is its recessive effect size. Therefore, treatment efficacy in each genetic subgroup and in the combined subgroup \{$Aa$, $aa$\} and \{$AA$, $Aa$\} are all of interest in order to identify the targeted patient population.

Application 2: Companion diagnostics studies that evaluate the threshold of a diagnostic marker (either continuous or ordinal with more than 2 categories). The decision to make is above (or below) which cutoff value of that marker, should the patients be diagnosed as ``suitable'' to get the drug (because of efficacy or safety). Patients with marker values on each side of the threshold are combinations of multiple marker subgroups.

In personalized medicine research, many existing and emerging biomarker and subgroup identification tools have to deal with mixture populations in RCTs. No matter it is an traditional subgroup analysis method that handles one biomarker at a time or it is an advanced recursive partitioning tree-based method that can handle many biomarkers simultaneously (for example, \cite{Foster2011} and \cite{Lipkovich2011}). And no matter the biomarkers are treated as categorical or continuous. Therefore, SME is a key component to all these biomarker methods and it has to be appropriately incorporated into these methods.

\bibliographystyle{biom}
\bibliography{biblio_YD}

\end{document}